\documentclass[prd, showpacs]{revtex4}
\usepackage{graphicx}
\usepackage{amssymb}
\usepackage{epsfig}
\usepackage{dcolumn}
\usepackage{bm}
\newcommand{\cc}{\cite}

\newcommand{\be}{\begin{equation}}
\newcommand{\ee}{\end{equation}}

\def\ve{\varepsilon}
\def\w{\omega}

\def\pd{\partial}
\def\f{\phi}
\def\F{\Phi}

\def\L{\Lambda}

\def\M{\bar\m}
\def\ex{\hbox{e}}

\def\F{\Phi}
\def\<{\langle}
\def\>{\rangle}

\def\a{\alpha}
\def\b{\beta}
\def\g{\gamma}  \def\G{\Gamma}
\def\d{\delta}  \def\D{\Delta}
\def\l{\lambda}   \def\L{\Lambda}
\def\s{\sigma}
\def\r{\rho}  
\def\x{\xi}
\def\c{\chi}

\def\m{\mu}
\def\n{\nu}

\def\w{\omega}

\def\v{\vec}

\def\vf{\varphi}
\def\({\left(}
\def\[{\left[}
\def\){\right)}
\def\]{\right]}
\def\coth{\hbox{coth}}

\def\cos{\hbox{cos}}

\def\pd{\partial}
\def\dk{{d^n k \over (2\pi)^n}}
\def\Tr{\hbox{Tr}}
\def\pa{{\cal P}}
\def\w1{W^{(1)}}
\def\v1{V^{(1)}}
\def\prop{D_{\m\n}}

\def\dI{\int \! dn(\r)}

\begin{document}
\title{\bf {Nonperturbative contributions
to the quark form factor at high energy}
\footnote{Extended version of
the talk presented at XXXI ITEP Winter School of Physics, 18--26 Feb 2003,
Moscow, Russia}}
\author{Igor O. Cherednikov}
\email{igorch@thsun1.jinr.ru}
\affiliation{{\sl Joint Institute for Nuclear Research \\
BLTP JINR, RU-141980 Dubna, Russia}}
\date{\today}
\begin{abstract}
The analysis of nonperturbative effects in high energy asymptotics of
the electomagnetic quark form factor is presented.
It is shown that the nonperturbative
effects determine the initial value for the
perturbative evolution of the quark form factor and find their general
structure with respect to the high energy asymptotics.
Within the Wilson integral formalism which is natural for investigation of the
soft, IR sensitive, part of the factorized form factor,
the structure of the instanton induced
effects in the evolution equation is discussed.
It is demonstrated that the instanton contributions result
in the finite renormalization
of the subleading perturbative result and numerically are characterized by small
factor reflecting the diluteness of the QCD vacuum within the instanton liquid
model. The relevance of the IR renormalon
induced effects in high energy asymptotic behaviour is discussed.
The consequences of the various analytization procedures of the strong coupling
constant in the IR domain are considered.
\end{abstract}
\maketitle
\section{\label{sec:level1}Introduction}
The electromagnetic (color singlet) quark form factor is one of the simplest and
convenient objects for the investigation of the double logarithmic
behaviour of the QCD amplitudes in the high energy regime.
From the methodological point of view, the consistent study
of such an asymptotic requires a perturbative
resummation procedure beyond the standard renormalization group techniques.
Besides this, the resummation methods developed for this
particular case can be applied to study of many other processes which possess
the logarithmic enhancements near the kinematic boundaries.
On the other hand, in addition to
the evident theoretical significance, the computation of the quark
form factors has important phenomenological applications.
The quark form factor enters into the
cross sections of a number of the high energy hadronic processes \cc{HARD}.
For example, the total cross section
of the Drell-Yan process (normalized to DIS) is determined by the ratio of the
time-like and space-like
form factors \cc{PAR, MAG}.
The similar resummation approach is used also in study of the near-forward
quark-quark scattering, and evaluation of the soft Pomeron properties \cc{KRP}. In the latter case, the nonleading
logarithmic terms are quite important. The investigation of the electomagnetic
quark form factors (Dirac as well as Pauli)
in moderate and low energy domains can shed a
light on the problem of the scaling violation in DIS and
the structure of constituent quarks \cc{SCAL}.

The first example of the large logarithm resummation was
given by Sudakov for the case of {\it off}-shell fermion in external Abelian
gauge field in the leading logarithmic approximation (LLA), where
the terms of order of $(\a_s^n  \ln^{2n} q^2)$ are taken into account while
the contributions from $O(\a_s^n \ln^{2n-1} q^2)$ are neglected.
The exponentiation of the leading
double logarithmic result was found \cc{SUD}. This
exponential decreasing of the form factor at large-$q^2$ means that the elastic scattering
of a quark by a virtual photon is suppressed at asymptotically large
momentum transfer.
The exponentiation for the {\it on}-shell form factor in the Abelian case was
obtained in the LLA in \cc{ONLLA}. As expected, the non-Abelian gauge theories
appeared to be more complicated: first,
the leading LLA terms in the QCD perturbative series
were found to be consistent with exponentiation in \cc{LLANA} (the inelastic
{\it on-}shell form factor
with emission of one and two gluons was calculated in the same
context in \cc{LLANAIE}; the role of the quark Sudakov form factor in the description
of $e^+e^-$ one-photon annihilation in quarks and gluons was considered in LLA in
\cc{FEL}), and the
all-order LLA non-Abelian exponentiation has been proved in \cc{LLAAOR}. In the
LLA, the exponentiated form factor was shown to be the rapidly decreasing
function at high momentum transfer, but the question if the non-leading
logarithmic terms could upset the LLA behaviour required a further work.
The all-(logarithmic)-order resummation was performed in the Abelian
case and the exponentiation was demonstrated in \cc{ALLLOGA}. In the paper
\cc{ALLLOGNA}, the non-Abelian all-order exponentiation for the so-called hard part of
the {\it on-}shell
form factor has been shown first within the powerful factorization approach.
Note, that in this work the case where a
time-like photon with large invariant mass decays into a quark-antiquark
pair was considered, however it can be easily shown that
the results remain true for our case of a quark scattering in an external
EM field as well.

In the work \cc{ALLLOGNA}, the detailed study of hard part
of the form factor (which is responsible for the UV properties) was
performed, while the status
of the soft part, containing all the IR and collinear
singularities and, as a consequence, all possible nonperturbative effects, remained unclear.
The important results on the IR properties of the QCD vertex functions was
obtained in \cc{KRC1} within the Wilson integral approach. In these works,
the soft part of the form factor had been presented as the vacuum averaged
ordered exponent of the path integral of a gauge field over the contour
of a special form---an angle with sides of infinite length. The use of the gauge
and renormalization group invariance allowed to derive the perturbative
evolution equation
describing the high energy behaviour of the form factor taking into
account all (not power suppressed) parts of the factorized amplitude,
both for the {\it on-} \cc{KRON} and
{\it off-}shell \cc{KROFF} cases. It was shown that the leading asymptotic
is controlled by the cusp anomalous dimension which arises due to the
multiplicative renormalization of the soft part, and can be calculated within
the Wilson integral formalism up to the two-loop order \cc{KRC1}.
It is worth noting that within the Wilson integral approach, the non-Abelian
exponentiation can be proved independently
\cc{NAEXP}, what is another important advantage of this framework.
The efficiency of the Wilson integrals approach had been successfully demonstrated
in a series of works \cc{STEF1, STEF2, STEF3}. In these papers, the
consistent non-diagrammatic framework
is developed what allows to calculate the fermionic Green's functions,
Sudakov form factors, amplitudes and cross sections in
QED and QCD completely in terms of world-line integrals, and thus avoid
complicated diagrammatic factorization analysis.

The results presented above allow one to conclude that the leading high
energy behaviour of the quark form factor in non-Abelian gauge theory is
completely determined by the perturbative
evolution equation, and is given by the fast decreasing exponent:
\be
\sim \exp\[- {2C_F \over \b_0}
\cdot \ln q^2 \ \ln \ln q^2 + O(\ln q^2)\]\ . \label{eq:LAS}
\ee This rapid fall off is not changed by any other logarithmic contributions
\cc{ALLLOGNA, KRON, KROFF, STEF2}.
However, the non-leading logarithmic corrections are nevertheless
important for evaluation of the numerical value of the form factor.
Some of them are of a purely
perturbative origin (higher loop corrections and sub-leading logarithmic
terms), while the others can be attributed to the nonperturbative phenomena.
The usual approach to treatment of the latter is developed within the IR renormalon
picture (there are plenty of papers on this subject, for the most recent
reviews see \cc{REN}). However they could only give the power-suppressed terms, which become,
of course, important in low energy domain, but can be neglected at
asymptotically large momenta. Here we should note, that this conclusion is
to be changed for processes with two scales (such as quark-quark scattering,
Drell-Yan process, {\it etc.}): then the corrections proportional to the
powers of a smaller scale must also be involved in the game
\cc{KRREN}. In the present work, we try to advocate the point of view
that the true (not connected directly to renormalons)
nonperturbative effects can be taken into account consistently in
the evolution equation, and therefore they yield the non-vanishing
subleading (perhaps, perimetrically suppressed, but still logarithmic)
contributions $\sim \ln \ q^2$ to the high energy behaviour. Further, we analyze
another possible source of contributions which can be considered
as ``nonperturbative''---the IR renormalon ambiguities.
We demonstrate explicitly that they produce the corrections with different
IR structure compared to that one generated by instantons. Moreover, as it can be shown
these direct renormalon effects disappear in the dimensional regularization \cc{MAG}
and in the analytical perturbation theory \cc{SHIR}, what means that one could hardly expect
a significant signature of the IR renormalon effects in this process.

The idea that the nontrivial vacuum structure could be relevant in high
energy hadronic processes was first explicitly formulated for the soft Pomeron
case in Abelian gauge theory by Low \cc{LOW}, Nussinov \cc{NUS}, and Landshoff and Nachtmann \cc{LN}, and
developed further using the eikonal approximation and the
Wilson integral formalism in \cc{NACH}.
In the present work, we consider the nonperturbative effects originating in
the nontrivial structure of the QCD vacuum treating the latter within the
framework of the instanton liquid model (ILM) \cc{ILM, REV, DP, DREV}. The approach based on the
other principles is successfully developed within the stochastic vacuum model (SVM),
where some important and interesting results have been obtained (see, {\it e.g.,} \cc{SVM}).
However, since the correspondence between both pictures is not completely
clear at the moment, we will restrict ourselves with the ILM only omitting
the discussions of relations with results of SVM.

The paper is organized as follows:
In Section II we describe the consequences of the RG invariance properties of the
factorized form factor, and
derive the linear evolution equation considering the
nonperturbative input as the initial value for perturbative
evolution. In Section III, the nonperturbative effects are estimated
in the weak-field approximation within the instanton model of the QCD vacuum.
The large-$q^2$ behaviour of the form factor is analyzed taking into account
the leading
perturbative and instanton induced contributions. In Section IV,
we study the consequences of the IR renormalon ambiguities of the perturbative series and
discuss their relevance within the context of certain analytization procedures.

\section{Evolution equation and nonperturbative effects}

The behaviour of the form factors in various energy
domains is one of the most important questions
in the theory of hadronic exclusive
processes. The electromagnetic quark form factors
are determined via the elastic scattering
amplitude of a quark in an external color singlet gauge field:
\be
{\cal M}_\m =F_q\[(p_1-p_2)^2\]  \bar u(p_1) \g_\m v(p_2)  -
G_q \[(p_1-p_2)^2\] \bar u(p_1) {\s_{\m\n} (p_1-p_2)_\n \over 2m} v(p_2)
 \ \ , \label{eq:ampl}
\ee where $u(p_1), \ v(p_2)$ are the spinors of outgoing and incoming
quarks, and $\s_{\m\n} = [\g_\m, \g_\n]/2$. In the high energy regime, the
Pauli form factor $G_q$ is power suppressed and will be neglected in
the present consideration. However, it should be emphasized that in low
and moderate energy domains it becomes important and interesting perturbative
as well nonperturbative effects arise (see, {\it e.g.,} recent works \cc{ET, KOCH}).

The kinematics of the process can described in terms of the scattering
angle $\c$:
\be \cosh \c = {(p_1 p_2) \over m^2} = 1 + {Q^2 \over 2 m^2} \ \ ,
\ \ Q^2 = - (p_2 - p_1)^2>0 \ \ , \ \ p_1^2=p_2^2 =m^2  \ .\label{m1} \ee
The classification of the diagrams with respect to the momenta carried by their
internal lines allows to express the form factor $F_q$ in the amplitude
(\ref{eq:ampl})
in the factorized form \cc{ALLLOGA, ALLLOGNA, KRON} (compare with the
world-line expression for the three-point vertex
in \cc{STEF2})
\be
F_q(q^2) = F_H(q^2/\m^2, \a_s)\cdot F_S(q^2/m^2, \m^2/\l^2, \a_s) \cdot F_J(\m^2/\l^2, \a_s)\
 , \label{eq:fact1}
\ee where the hard, soft, and collinear (jet) part are separated. Note, that in the present paper,
all the dimensional variables are assumed to be expressed in units of the QCD scale $\L$, so that
$q^2 = Q^2/\L^2$, {\it etc}. The arbitrary scale $\m^2$ stands for the boundary value of the
squared internal momenta which divides the different parts, and is assumed to be
equal to the UV normalization point.

The total form factor $F_q$ is the renormalization invariant quantity:
\be
\m^2 {d \over d \m^2} \ F_q (\m^2, \a_s(\m^2))= \(\m^2 {\pd \over \pd \m^2} +
\b(\a_s){\pd \over \pd \a_s}\) \ F_q (\m^2, \a_s(\m^2)) =0 \ ,
\ee what leads, in the large-$q^2$ regime, to the following relations
\be
\m^2 {d \over d \m^2}\[ {\pd \ln F_H \over \pd \ln q^2}\] = - \m^2 {d \over d \m^2}
\[{\pd \ln F_S \over \pd \ln q^2} \]=
{1 \over 2}\ \G_{cusp} (\a_s) \ . \label{eq:anom}
\ee For convenience, we work with the logarithmic derivatives in $q^2$ what
allows us to avoid the problems with additional light-cone singularities in the soft part
\cc{KRON, KRLC}. The collinear part $F_J$ being independent on $q^2$ does not contribute
to these equations.

Within the eikonal approximation, the resummation of all logarithmic terms
coming from the soft gluon subprocesses
allows us to express $F_S$ in terms of the vacuum average of the gauge invariant path ordered
Wilson integral \cc{MMP}
\be
F_S (q^2/m^2, \m^2/\l^2, \a_s) = W (C_\c; \m^2/\l^2, \a_s)  =
{1 \over N_c} \Tr \<0|  \pa \exp \left\{ i g \int_{C_\c} \! d x_{\m} \hat A_{\m}
(x) \right\}|0\> \ . \label{1a} \ee
In Eq. (\ref{1a}) the integration path corresponding to considering process
goes along the closed contour $C_\c$: the angle (cusp) with infinite sides.
The gauge field $
\hat A_{\m} (x) = T^a A^a_{\m}(x)\ , \   \ \Tr[T^a T^b] = \frac{1}{2}\delta^{ab}, \ $
belongs to the Lie algebra of the gauge group
$SU(N_c)$, while the Wilson loop operator $\pa \exp\({ig\int\! dx A(x)}\)$ lies in
its fundamental representation.
The cusp anomalous dimension $\G_{cusp}$ can be found from the multiplicative renormalization
of the Wilson integral (\ref{1a}) \cc{KRC1, WREN}:
\be
W(C_\c; \m^2/\l^2, \a_s(\m^2)) = Z_{cusp} (C_\c; \M^2/\m^2, \a_s(\m^2)) \cdot W_{bare} (C_\c; \M^2/\l^2, \a_s)
\ , \label{eq:renw1}
\ee where $\M^2$ is the UV cutoff, $\m^2$ is the normalization point, and $\l^2$ is the IR cutoff.
The presence of the IR divergence in (\ref{eq:renw1}) is a common feature of {\it on}-shell
amplitudes in massless QCD.
Since $W_{bare}$ knows nothing about the normalization point (the latter is
fixed by choosing a concrete $Z_{cusp}$), one can write:
\begin{eqnarray}
{1 \over 2}\ \G_{cusp} (C_\c; \a_s(\m^2)) =
- \m^2 {d \over d \m^2} \ln W(C_\c; \m^2/\l^2, \a_s(\m^2)) =
- \m^2 {d \over d \m^2} \ln Z_{cusp}(C_\c; \M^2/\m^2, \a_s(\m^2) ) \ . \label{eq:renw2}
\end{eqnarray} It can be shown that the cusp anomalous dimension (\ref{eq:anom}) is linear in
$\c$ to all orders of perturbation theory in the large-$q^2$ regime \cc{KRC1}:
\be
\G_{cusp} (C_\c; \a_s) = \ln q^2 \ \G_{cusp} (\a_s) + O(\ln^0 q^2)\ . \label{eq:cad}
\ee
Then, from the Eqs. (\ref{eq:anom}, \ref{eq:renw2}, \ref{eq:cad}) one finds after the simple calculations
\cc{KRON}:
\be
{\pd \ln F_H (q^2) \over \pd \ln q^2} = \int_{q^2}^{\m^2} \! {d\x \over 2\x} \G_{cusp} (\a_s(\x)) + \G(\a_s(q^2)) \ ,
\label{eq:solH}
\ee
\be
{\pd \ln F_S (q^2) \over \pd \ln q^2} = - \int_{\l^2}^{\m^2} \! {d\x \over 2\x} \G_{cusp} (\a_s(\x)) +
{\pd \ln W_{np} (q^2) \over \pd \ln q^2} \ ,
\label{eq:solS}
\ee where the ``integration constant'' of the hard part reads
\be
\G(\a_s(q^2)) = {\pd \ln F_H (q^2)\over \pd \ln q^2}\Bigg|_{\m^2 = q^2} \ , \label{eq:intch}
\ee
and $W_{np}$ arises as the initial value of the soft part:
\be
{\pd \ln W_{np} (q^2) \over \pd \ln q^2}  = {\pd \ln F_S(q^2) \over \pd \ln q^2}
\Bigg|_{\m^2 = \l^2} \ , \label{eq:intcs}
\ee and is the only quantity where, according to our suggestion, the nonperturbative
effects could take place \cc{DCH1, DCH2}.
Then we get the $q^2$-evolution equation of the total form factor at large $q^2$:
\begin{eqnarray}
&&\ln {F_q(q^2) \over F_q(q^2_0)}
= - \int_{q_0^2}^{q^2}\! {d\x \over 2\x} \ \[ \ln{q^2 \over \x}
\ \G_{cusp}(\a_s(\x) ) - 2 \G(\a_s(\x))\] - \ln {q^2 \over q_0^2}\ \int_{\l^2}^{q_0^2} \! {d\x \over
2\x} \ \G_{cusp}(\a_s(\x) ) +  \ln {W_{np} (q^2) \over W_{np} (q_0^2)} \ . \label{eq:gen1}
\end{eqnarray}
In the next Section, we calculate explicitly the perturbative quantities entering Eq. (\ref{eq:gen1})
in one loop approximation.

\section{Analysis of the perturbative contributions to the Wilson integral}

The analysis of the hard contributions
\cc{ALLLOGNA, KRON} at large $q^2$ yields:
\be
{\pd \ln F_H(q^2/\m^2, \a_s) \over \pd \ln q^2} = - {\a_s \over 2 \pi} C_F \(\ln {q^2 \over \m^2}
- {3 \over 2} \) + O(\a_s^2)\ , \label{eq:hard1}
\ee where $C_F = (N_c^2 -1)/2N_c$.
For the hard ``integration constant'' (\ref{eq:intch}) one has:
\be
\G(\a_s(q^2)) = {3 \over 4} {\a_s(q^2) \over \pi} C_F \ .
\ee
The expression (\ref{eq:hard1}) is IR-safe, while
the low-energy information is accumulated in the soft part of the quark form factor $F_S$.
The Wilson integral (\ref{1a}) can be presented as a series:
\begin{eqnarray}
W(C_\c) && = 1 + {1 \over N_c}\<0| \sum_{n=2} \ (ig)^n \int_{C_\c}\int_{C_\c} ...\int_{C_\c}
 \! dx_{\m_n}^n \ dx_{\m_{n-1}}^{n-1}... dx_{\m_1}^1 \cdot\nonumber\\ &&
\cdot \ \theta (x^n, x^{n-1}, ... , x^1)
\ \Tr \[\hat A_{\m_n} (x^n) \hat A_{\m_{n-1}} (x^{n-1})... \hat A_{\m_{1}} (x^1)
\]|0\> \ , \label{expan1} \end{eqnarray} where the function $\theta (x)$ orders the color matrices along
the integration contour.
In the present work, we restrict ourselves with the study of the  leading order
(one-loop  --- for the perturbative gauge field and weak-field limit for the
instanton) terms of the expansion (\ref{expan1}) which are given by the
expression:
\be
\w1_{bare} (C_\c) =  -{g^2 C_F \over 2}
\ \int_{C_\c}\! dx_\m \int_{C_\c}\! dy_\n \ \prop (x-y)
\ , \label{g1} \ee where the gauge field propagator $\prop(z)$ in
$n$-dimensional space-time $(n = 4 - 2 \ve)$ can be presented in the form:
\be
\prop(z) = g_{\m\n}
\pd_z^2 \D_1(\ve, z^2, \M^2/\l^2) - \pd_\m\pd_\n \D_2(\ve, z^2, \M^2/\l^2) \ .
\label{st1} \ee Here $\M^2$ is the parameter of the dimensional regularization.
The exponentiation theorem for non-Abelian path-ordered Wilson integrals
\cc{NAEXP} allows us to express (to one-loop accuracy) the Wilson integral (\ref{1a})
as the exponentiated one-loop term of the series (\ref{expan1}):
\be
W_{bare}(C_\c; \ve, \M^2/\l^2) = \exp\[\w1_{bare}(C_\c; \ve, \M^2/\l^2) + O(\a_s^2)\] \ . \label{eq:expn1}
\ee
In general, the expression (\ref{g1}) contains ultraviolet (UV) and IR divergences, that
can be multiplicatively renormalized in a consistent way \cc{WREN}.
In the present work, we use the dimensional regularization for the UV singularities,
and define the ``gluon mass'' $\l^2$ as the IR regulator.
The dimensionally regulated formula for the leading order (LO) term (\ref{g1})
can be written as \cc{DCH1}:
\be
\w1_{bare} (C_\c; \ve, \M^2/\l^2, \a_s)
= 8 \pi \a_s C_F h (\c) (1 - \ve)  \D_1(\ve, 0, \M^2/\l^2) \ ,
\label{pe1} \ee
where $h(\c)$ is the universal cusp factor:
\be
h(\c) = \c \coth \c -1 \  \ee
and for perturbative gauge field
\be
\D_1 (\ve, 0, \M^2/\l^2) = - {1 \over 16\pi^2} \(4\pi {\M^2 \over
\l^2}\)^{\ve} \ {\G(\ve) \over 1 - \ve} \ .  \label{eq:pr01}
\ee
The independence of the expression (\ref{pe1}) from the function $\D_2$ is a
direct consequence of the gauge invariance. Then, in the one-loop approximation,
\be
W_{bare}(C_\c; \ve, \M^2/\l^2, \a_s) = 1 - {\a_s \over 2\pi} C_F h(\c) \({1 \over \ve}
- \g_E + \ln 4\pi + \ln{\M^2\over\l^2}
\),
\ee and the cusp dependent renormalization constant, within the
$\overline{MS}$-scheme which fixes the UV normalization point, reads:
\be
Z_{cusp} (C_\c; \ve, \M^2/\m^2, \a_s(\m^2)) =
1 + {\a_s(\m^2) \over 2\pi} C_F h(\c) \({1 \over \ve} - \g_E + \ln 4\pi \) +O(\a_s^2)\
. \ee

Using the Eq. (\ref{pe1}), one finds the known one-loop result
for the perturbative field, which  contains the dependence
on the UV normalization point $\m^2$ and IR cutoff $\l^2$
({\it e.g.}, \cc{KRC1, STEF2}):
\be \w1_{pt} (C_\c; \m^2/\l^2, \a_s(\m^2))
=  - {\a_s (\m^2) \over 2 \pi} C_F  h(\c) \ln {\m^2 \over \l^2} + O(\a_s^2)
\ . \label{5} \ee
Therefore, in the leading order the kinematic dependence of the expression (\ref{g1}) is
factorized into the function $h(\c)$, which at large-$q^2$ is approximated by:
\be h(\c) \propto \ln {q^2 \over m^2} \ . \label{eq:lar1} \ee
From the one-loop result (\ref{5}),
the cusp anomalous dimension which satisfies the RG equation
(\ref{eq:renw2}) in one-loop order is given by:
\be {\G_{cusp}^{(1)}} ( \a_s (\m^2)) = {\a_s (\m^2) \over \pi} C_F \ .  \label{cp} \ee
Substituting the anomalous dimension (\ref{cp}) in the one-loop approximation for the strong
coupling into the Eq. (\ref{eq:gen1}), one finds
\begin{eqnarray}
&& F_q^{(1)}(q^2) = \exp\[- {2C_F \over \b_0} \[\ln q^2 \(\ln{\ln q^2 \over \ln \l^2} -1 \)
- {3 \over 2} \ln {\ln q^2 \over \ln q_0^2 }
+ \ln q_0^2
\(1 - \ln{\ln q_0^2 \over \ln \l^2} \)  \] +  W_{np}(q^2) \]
F^{(1)}(q_0^2)\ . \label{npc0} \end{eqnarray}
Note, that the exponent in Eq. (\ref{npc0}) has an unphysical singularity at
$\l^2 =1$ (in dimensional notations, $\bar\l^2 = \L_{QCD}^2$), {\it
i. e., } where the coupling constant $\a_s (\bar\l^2)$ has the Landau pole. This feature
can be treated in terms of the IR renormalon ambiguities (see Section V), and is considered often
as a signal of nonperturbative physics. In the present paper, we will consistently separate the
sources of nonperturbative effects which can be attributed to uncertainties of the perturbative series
resummation, from the ``true'' nonperturbative phenomena. An important example of the latter is
provided by the instanton induced effects within the ILM of QCD vacuum, which is considered
in the next Section.

\section{Large-$q^2$ behaviour of the instanton induced contribution}

The instanton induced effects in the high energy QCD processes
had been studied actively since the seventies \cc{EL, BAL}).
Recently, the investigation of these effects renewed from the fresh points of view
\cc{Sh, KKL, RW, SRP, RING, DCH1, DCH2, D03, KOCH}. The Wilson integral formalism is considered as a useful
and convenient tool in the instanton calculations, mainly due to the significant simplification
of the path integral evaluation for an explicitly known gauge field \cc{Sh}. Another important
feature of this approach is the possibility of a correct analytical continuation of the results obtained
in the Euclidean picture (where the instantons are only determined) to the realistic Minkowski space-time,
where the scattering process (\ref{eq:ampl}) actually takes place \cc{EUC}.
Namely, in the instanton calculations, one maps
the scattering angle, $\c$, to the Euclidean space by the analytical continuation
\be \c \to i\g \ , \ee and performs the inverse transition to the Minkowski space-time in the final
expressions in order to restore the $q^2$-dependence.
Let us consider the instanton induced contribution to the function $W_{np}(q^2)$ from Eq.(\ref{eq:intcs}).
The instanton field is given by
\be \hat A_\m (x; \r) = A^a_{\m} (x; \r) {\sigma^a \over 2} = {1 \over g}
 {\hbox{\bf R}}^{ab} \sigma^a {\eta^\pm}^b_{\m\n} (x-z_0)_\n \vf
(x-z_0; \r) , \label{if1}\ee
where ${\hbox{\bf R}}^{ab}$ is the colour orientation matrix $(a = {1,..., (N_c^2-1)}, b=1,2,3)$ which
provides an embedding of $SU(2)$ instanton field into $SU(3)$ colour group,
$\s^a$'s are the Pauli matrices,
and $(\pm)$ corresponds to the instanton, or anti-instanton.
The averaging of the Wilson operator over the nonperturbative vacuum is performed by
the integration over the coordinate of the instanton center $z_0$, the color orientation and the
instanton size $\r$.  The measure for the averaging over the instanton ensemble
reads $dI = d{\hbox{\bf R}} \ d^4 z_0 \ dn(\r) $, where
$ d{\hbox{\bf R}}$ refers to the averaging over color orientation,
and $dn(\r)$ depends on the choice of the instanton size distribution.
Taking into account (\ref{if1}), we write the Wilson integral (\ref{1a})
in the single instanton approximation in the form:
\be
w_I(C_\g) = {1\over N_c}  \<0| \Tr  \exp \( i \sigma^a \phi^a \)|0\> \ ,
\label{wI1}\ee
where
\be \phi^a(z_0,\r) =
{\hbox{\bf R}}^{ab} {\eta^\pm}^b_{\m\n} \int_{C_\g} \! dx_\m \ (x-z_0)_\n
\vf (x-z_0; \r) \ . \label{iin} \ee We omit the path
ordering operator $\pa$ in (\ref{wI1}) because the instanton field
(\ref{if1}) is a hedgehog in color space, and so it locks the color
orientation by space coordinates.
One obtains the all-order single instanton contribution in the form \cc{DCH1}:
\be
w_I(C_\g) =   \int \! d^4 z_0 \int \! dn(\r) \  \[\cos \ \phi (\g, z_0, \r) - \cos \ \phi (0, z_0, \r)  \]\
, \label{it1}
\ee
where the squared phase $\phi^2 = \phi^a\phi^a$ may be written as
\begin{eqnarray}
\phi^2 (\g, z_0, \r) = && \nonumber\\ = \sum_{i,j=1,2} \[ (v_iv_j) z_0^2   -
(v_iz_0)(v_jz_0)\]
\int_0^\infty \! d\s \vf\[((-1)^{i+1} v_i\s - z_0)^2; \r \] &&
\int_0^\infty \! d\s' \vf \[((-1)^{j+1} v_j\s' - z_0)^2; \r \]  \ ,  \label{it2}
\end{eqnarray} where $v_{1,2} = p_{1,2}/m $, and $v_{1,2}^2 = 1 \ , \ (v_1v_2) = \cos \g$ in Euclidean geometry.
Although sometimes such integrals (Eq. (\ref{it2})) can be evaluated
explicitly, the given contour requires numerical calculations,
thus we restrict ourselves with the weak-field approximation which can be
studied analytically.
Then, in case of the instanton field, the LO contribution in Minkowski space reads
\be
w_I^{(1)}(C_\c) = 2 h(\c) \int\! dn(\r) \ \D_1^I(0, \r^2\l^2) \ ,
\ee where
\be
\D_1^I(0, \r^2\l^2) = -  \int \! {d^4 k \over (2\pi)^4} \ex^{ikz}
\d(z^2) \[2 \tilde \vf'(k^2; \r)\]^2\ ,  \ee
and we use the same IR cutoff $\l^2$, while the UV
divergences do not appear at all due to the finite instanton size.
Here, $\tilde \vf (k^2; \r)$
is the Fourier transform of the instanton profile
function $\vf (z^2; \r)$ and $\tilde \vf'(k^2; \r)$ is it's derivative with respect to
$k^2$.
In the singular gauge, when the profile function is
\be
\vf (z^2) = {\r^2 \over z^2 (z^2 + \r^2)} \ ,
\ee one gets:
\be
\D_1^I (0, \r^2\l^2) = {\pi^2 \r^4 \over 4} \[\ln (\r^2 \l^2) \ \F_0(\r^2\l^2) + \F_1(\r^2\l^2)
\]\ ,
\ee where
\be
\F_0 (\r^2\l^2) = {1 \over \r^4\l^4} \int_0^1\! {dz \over z (1-z)} \ \[1+ \ex^{\r^2\l^2} -
2 \ex^{z \cdot \r^2\l^2} \]
\ \ , \ \ \lim_{\l^2\to 0}\F_0(\r^2\l^2) = 1 \ ,
\ee and
\be
\F_1(\r^2\l^2) = \sum_{n=1}\int_0^1 \! dxdydz \ {[-\r^2\l^2 (xz + y(1-z))]^n \over n!
n}\ex^{\r^2\l^2 [xz + y(1-z)] } \ \ , \ \ \lim_{\l^2\to 0}\F_1(\r^2\l^2) = 0
\
\ee are the IR-finite expressions. At high energy the instanton induced contribution
is reduced to the form:
\be
{\pd  \ln W_I(q^2) \over \pd  \ln q^2} = {\pi^2 \over 2}
\dI \ \r^4 \[\ln (\r^2 \l^2) \ \F_0(\r^2\l^2) + \F_1(\r^2\l^2)
\] \equiv - B_I(\l^2)\ .  \label{II1}\ee
Here we used the exponentiation of the single-instanton result
in a dilute instanton ensemble \cc{DCH1}:  \be W_I = \exp\(w_I\)
\ , \label{eq:exp} \ee and took only the LO term of the weak-field
expansion (\ref{g1}): $\w1 = w_I + (higher \ order \ terms)$.

In order to estimate the magnitude of the instanton induced effect
we consider the standard  distribution function \cc{tH} supplied with the
exponential suppressing factor, what has been suggested in
\cc{SH2} (and discussed in \cc{DEMM99} in the framework of constrained instanton
model) in order to describe the lattice data \cc{LAT}:
\be
dn(\r) = {d\r \over \r^5} \ C_{N_c} \[2\pi \over \a_s(\m_r) \]^{2N_c} \exp\[- {2\pi \over \a_s (\m_r)}
\] \(\r\m_r\)^{\b} \exp\(- 2 \pi \s \r^2\) \ , \label{dist1}
\ee
where the constant $C_{N_c} = 4.6/\pi^2 \ \exp(-1.679 N_c)/\[(N_c-1)! (N_c-2)!\] \approx 0.0015 $,
$\s$ is the string tension,  $\b= \b_0+O(\a_s(\m_r))$, and $\m_r$ is
the normalization point \cc{MOR}. Given the distribution (\ref{dist1})
the main parameters of the instanton liquid model---the instanton density $\bar n$
and the mean instanton size $ \bar \r $---will read:
\be
\bar n = \int_0^\infty \! dn(\r) = {C_{N_{c}} \G (\b/2 - 2) \over 2} \[2\pi \over \a_S(\bar \r^{-1}) \]^{2N_c}
\[{\L_{QCD} \over \sqrt{2\pi \s}}\]^{\b} (2\pi \s)^2 \ ,
\ee
\be
\bar  \r = {\int_0^\infty \! \r \ dn(\r) \over \int_0^\infty \! dn(\r)} = {\G(\b/2 - 3/2) \over \G(\b/2 - 2)} {1 \over \sqrt{2 \pi \s} } \ . \label{nbar} \ee
In Eq. (\ref{nbar}) we choose, for convenience, the normalization scale
$\m_r$ of order of the instanton inverse mean size $\bar\r^{-1}$, taking into account
that the distribution function (\ref{dist1}) in the RG-invariant quantity up to
$O(\a_s^2)$ terms \cc{MOR}.
Note, that these quantities correspond to the mean size $\r_0$ and
density $n_0$ of instantons used in the model \cc{ILM}, where the size distribution
(\ref{dist1}) is approximated by the delta-function:
$
dn(\r) = n_0 \d(\r-\r_0) d\r \ .$

Thus, we find the leading instanton contribution (\ref{II1}) in the form:
\be
B_I=  K \pi^2 \bar n {\bar \rho}^4 \ln{2\pi \s \over \l^2} \[1 + O\({\l^2\over 2\pi\s}\)\]
, \label{pow1} \ee
where
\be
K = {\G(\b_0/2) [\G (\b_0/2-2)]^3 \over 2 \ [\G(\b_0/2-3/2)]^4} \approx
0.74 \ ,
\ee
and we used the one loop expression for the running coupling constant
\be
\a_s (\bar \r^{-1}) = - {2\pi \over \b_0 \ln \ {\bar \r \L}} \ \ ,
 \ \ \b_0 = {11 N_c -
2 n_f \over 3}\ \ . \ee
The packing
fraction $ \pi^2 \bar n {\bar \rho}^4 $ characterizes diluteness
of the instanton liquid and within the conventional picture its value is estimated to be
$ 0.12 \ ,$
if one takes the model parameters as (see \cc{REV}):
\be  {\bar n} \approx 1 fm^{-4},
\ \ {\bar \r} \approx 1/3 fm \ ,
\ \ \s \approx (0.44 \ GeV)^2. \label{param} \ee
The leading logarithmic contribution to the quark form factor at asymptotically large $q^2$
is provided by the (perturbative) evolution governed by
the cusp anomalous dimension (\ref{cp}). Thus, the instantons yield the
subleading effects to the large-$q^2$ behaviour accompanied by a numerically small
factor as compared to the perturbative term:
\be B_I   \approx 0.02 << \frac{2C_F}{\beta_0} \approx0.24 \ . \ee

\begin{figure*}
\centering
\epsfig{file=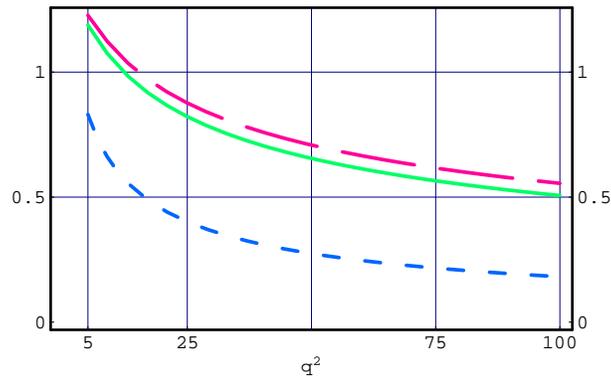,width=0.45\hsize}
\caption{The asymptotic behaviour
of the quark form factor is shown as the function of
the dimensionless variable $q^2 = Q^2/\L^2$, up to terms $O(\ln \ln q^2)$.
The long-dashed (red) presents the contribution of one loop
perturbative terms; the solid (green) line represents the total
form factor including the instanton induced part, Eq. (\ref{eq:final}).
For comparison, the leading $(\sim \ln q^2 \ln \ln q^2)$ perturbative contribution
is shown separately---the short-dashed (blue) line.}
\label{fig1}
\end{figure*}

Therefore, from Eqs. (\ref{II1}) and (\ref{npc}),
we find the expression for the quark form factor at large-$q^2$
with the one-loop perturbative contribution and the nonperturbative contributions
(the function $W_{np}$ in Eq. (\ref{npc0})) which include both the instanton induced terms:
\be
F_q(q^2) =  \exp\[- {2C_F \over \b_0} \ln q^2 \ln \ln q^2  -
\ln q^2 \ \(B_I - {2C_F \over \b_0} \) + O(\ln \ln q^2) \] F_0 (q_0^2; \l^2)
\ .  \label{eq:final} \ee
It is clear, that while the asymptotic ``double-logarithmic'' behaviour is
controlled by the
perturbative cusp anomalous dimension, the leading nonperturbative corrections
result in a finite renormalization of the subleading perturbative term
(Fig.\ref{fig1}). Note, that the instanton correction has the opposite sign compared
to the perturbative logarithmic term.

\section{Ambiguities of the perturbative result: IR renormalons and analytization of the coupling constant}

As it was pointed out in the end of Section II, the perturbative
evolution equation (\ref{npc0})
possesses an unphysical singularity at the point $\l^2 = 1$.
Therefore, it is instructive to study the consequences of this feature.
It is known that the presence of the Landau pole in the (one-loop)
expression for the coupling constant leads to
the IR renormalon poles in the Borel plane \cc{REN}
which result in the renormalon-induced power corrections.
The latter, being beyond the perturbative evaluations, are treated as
a signal of nonperturbative effects which should compensate the ambiguities
arising due to these poles.
In the present situation, one can expect
the corrections proportional to the powers of both scales: $\m^2$ and
$\l^2$. We assume here that the power $\m^2$-terms are
too strongly suppressed in large-$q^2$ regime and thus can be neglected in the
given context,
and focus on the power $\l^2$-corrections.
To find them, let us consider the perturbative function $\D_1(\ve, 0, \M^2/\l^2)$ in the Eq. (\ref{pe1}).
The insertion of the fermion bubble 1-chain to the one-loop order expression
(\ref{g1}) is equivalent to replacement of the frozen coupling constant $g^2$ by the
running one $g^2 \to g^2 (k^2) = 4\pi \a_s(k^2)$
\cc{KRREN} (for convenience, we work here in Euclidean space):
\be \widetilde\D_1(\ve, 0, \M^2/\l^2) = - 4\pi \M^{2\ve}  \int \! \dk \a_s(k^2){\ex^{ikz} \delta(z^2)
 \over k^2(k^2+\l^2)}\ .
\label{ren1} \ee
Using the integral representation for the one-loop running coupling $\a_s(k^2) = \int_0^\infty \!
d\s (1/k^2)^{\s b }$, $b = \b_0 /4\pi$, we find:
\be
\widetilde\D_1(\ve, 0, \M^2/\l^2) = - {1 \over \b_0(1-\ve)} \(4\pi {\M^2 \over \l^2}\)^\ve \int_0^{\infty} \!
dx\
{\G(1-x -\ve) \G(1+x+\ve) \over (x +\ve ) \G(1 - \ve )} \({1 \over \l^2}\)^x \ . \label{gamma} \ee
To define properly the integral in
r. h. s. of Eq.(\ref{gamma}), one needs to specify a prescription to go around the poles, which
are at the points $\bar x_n = n , \ n \in {\mathbb N}$. Thus, the result
of integration depends on this prescription giving an ambiguity proportional
to $\(1 / \l^2 \)^n$ for each pole. Then, the IR renormalons produce the power
corrections to the one-loop perturbative result, which we assume to exponentiate with the
latter \cc{KRREN}. Extracting from (\ref{gamma}) the UV singular part in vicinity of the
origin $x =0$, we divide the integration interval $[0, \infty]$ in two parts:
$[0, \d]$ and $[\d, \infty]$, where $\d < 1$. This procedure allows us to
evaluate separately the ultraviolet and the renormalon-induced pieces. For the
ultraviolet piece, we apply the expansion of the integrand in
$\D_1$ in powers of small $x$ and replace the ratio of $\G$-functions by
$\exp(-\g_E \ve)$:
\be \widetilde\D_1^{UV} (\ve, 0, \M^2/\l^2) = -  {1 \over \b_0 (1- \ve)}
\sum_{k, n=0} (-)^n { \(  \ln 4\pi - \g_E + \ln {\M^2  \over \l^2}\)^k
\over k! \ve^{n-k+1}}  \
\int_0^{\d} \! dx \ x^{n} \ \(1 \over \l^2 \)^x \ , \label{rnm1} \ee
which after subtraction of the poles in the $\overline{MS}$-scheme becomes:
\be
\widetilde\D_1^{UV} (0, \m^2/\l^2) =  {1 \over \b_0 ( 1- \ve)}\  \sum_{n=1}\(\ln {\m^2 \over \l^2}
\)^n {(-)^n \over n!} \ \int_0^{\d} \! dx x^{n-1} \(1 \over \l^2 \)^x \ .
\label{eq:iks}\ee
In analogy with results of \cite{Mikh98}, this expression may be rewritten in a closed form as
\be
\widetilde\D_1^{UV} ( 0, \m^2/\l^2) =
{1 \over \b_0 (1- \ve)} \int_{0}^{\d}\frac{dx}{x} \[\ex^{-x \ln \m^2}
-  \ex^{-x\ln \l^2} \].
\label{Dex}\ee
Then, using the relation
\be
{\pd  \w1 (q^2) \over \pd \ln q^2}= 2 C_F (1-\ve) \widetilde \D_1^{UV}(0, \m^2/\l^2)
\ , \label{Dex1}\ee
one finds
\be
\(\m^2 {\pd \over \pd \m^2} + \b(g) {\pd \over \pd g} \)
{\pd \ln W^{(1)} (q^2) \over \pd \ln q^2}
 = - {1 \over 2} \G_{cusp}^{(1)} (\a_s(\m^2)) \(1-\exp{\[-\d\frac{4\pi}{\b_0\a_s(\m^2)}\]}\).
\ee
The second exponent in the last equation yields the power suppressed
terms $\(1/q^2\)^\d$ in large-$q^2$ regime.
In the leading logarithmic approximation (LLA) Eq. (\ref{Dex1}) is reduced to:
\be
{\pd \w1 (q^2) \over \pd \ln q^2} = - {2C_F \over \b_0} \(
\ln{\ln \m^2 \over \ln \l^2}  \) \ . \label{ptr2}
\ee
The last expression obviously satisfies the perturbative evolution
equation (\ref{npc0}).

The remaining integral in Eq. (\ref{gamma}) over the interval $[\d, \infty]$
is evaluated at $\ve =0$ since there are no UV singularities.
The resulting expression does not depend on the normalization point $\m$, and thus
it is determined by
the IR region including nonperturbative effects. It contains the renormalon ambiguities
due to different prescriptions in going around the poles $\bar x_n$ in the Borel
plane which yields the power corrections to the quark form factor.

After the substitution $\m^2 = q^2$ and integration, we find
in LLA (for comparison, see Eq. (\ref{npc0})):
\be
F_q^{ren}(q^2) = \exp\[- {2C_F \over \b_0} \ln q^2
\(\ln \ln q^2  - 1 \) -  \ln q^2 \Phi_{ren} (\l^2) \] F^{ren}(q_0^2)\ , \label{npc}
\ee
where the function $ \Phi_{ren}(\l^2) = \sum_{k=0} \f_k (1/\l^2)^k$
accumulates the effects of the IR renormalons. The coefficients
$\f_k$ cannot be calculated in perturbation theory and are treated often
as ``the minimal set'' of nonperturbative parameters.  It is worth noting that
the logarithmic $q^2$-dependence of the renormalon induced corrections in the large-$q^2$
regime is factorized, and thus the Eq. (\ref{npc}) corresponds
to the structure of nonperturbative contributions found in the
one-loop evolution equation (\ref{npc0}), in a sense of its large-$q^2$ behaviour.
On the other hand,
the IR structures of the renormalon corrections
and the instanton induced ones (\ref{II1}) are different.
In our point of view, it allows us to separate the
true nonperturbative ({\it e.g.,} instanton induced, but not only)
effects from that ones related to ambiguities of the resumed perturbative series.

The latter conclusion can be illustrated by considering the consequences of an analytization
of the strong coupling constant \cc{SHIR} in the perturbative evolution equation.
In this approach,
the one-loop strong coupling $\a_s(\m^2)$ is replaced by the expression which is analytical
at $\m^2 = 1$ ({\it i.e.,} at $\L_{QCD}$ in dimensional variables):
\be
\a_s^{AN}(\m^2) = {4\pi \over \b_0} \( {1 \over \ln \m^2} + {1 \over 1 - \m^2} \) \ . \label{eq:an1}
\ee
The direct substitution of (\ref{eq:an1}) into the evolution equation (\ref{eq:gen1}) yields (for brevity,
we assume $q_0^2 = \l^2$):
\begin{eqnarray}
&& - \({\b_0 \over 2C_F} \) \ln F_q^{AN} (q^2) =  \nonumber\\ && =
\ln q^2 \ln{\ln q^2 \over \ln \l^2} - \ln {q^2 \over \l^2} -
{3 \over 2} \ln {\ln q^2 \over \ln \l^2 } +  \nonumber\\
&& + \ln q^2 \(\ln {q^2 \over q^2 -1} + \ln {\l^2 -1 \over \l^2} \)
- {1 \over 2} \( \ln^2 q^2 - \ln^2 \l^2 \) - \nonumber\\ && -
\hbox{Li}_2 (1-q^2) + \hbox{Li}_2 (1-\l^2)  + {3 \over 2 }
\( \ln {q^2 \over q^2 -1} + \ln {\l^2 -1 \over \l^2} \) \ . \end{eqnarray}
The functions $\hbox{Li}_2$ in the resulting expression
accumulate the power corrections of $q^2$ as well as IR scale $\l^2$, but
does not exhibit a singularity at $\l^2=1$.
Therefore, it gives no room for IR renormalons ambiguities, at least in
the considered approximation. Nevertheless, the power corrections of
a nonperturbative origin do contribute to the large-$q^2$ behaviour, and
the investigation of the correspondence between latter and the instanton corrections
calculated in the previous Section would be an interesting task.
Note, that the consequences
of the analytization of the strong coupling constant in the IR region have been
studied earlier in the case of the Sudakov effects in the pion form factor and Drell-Yan
cross section in the works \cc{STAN}.

Another possible way to avoid the Landau pole on the integration path
have been developed within the dimensional regularization \cc{MAG}.
In this case, the running coupling reads
\be
\a_s^{DR} (\ve; \m^2) =  {4 \pi \ve \over \b_0 \[ \(q^2 \)^\ve -1 \]} \ , \label{eq:dr1}
\ee and for complex $\ve \ ,\  \hbox{Re}\  \ve < 0$ it has the
Landau pole at the complex value of $\m^2$, that is this singularity appears to be
out of the integration contour. In the limit $\ve \to 0$, the form factor reads \cc{MAG} (for comparison, see
Eq. (\ref{npc0}):
\be F_q^{DR} (q^2) = \exp\[- {2 C_F \over \b_0} \({\zeta(2) \over \ve} + \ln \ve \(\ln q^2 - {3 \over 2}\)
+ \ln q^2 \(\ln \ln q^2 -1 \) - {3 \over 2} \ln \ln q^2 \) + O(\ve, \ve \ln \ve) \] \ .
\ee This expression also leaves no room for any renormalon induced effects.
In the same time, the instanton induced contribution still
takes place since they enter into the ``integration constant'' $W_{np}$ which knows nothing about the analytical
properties of the coupling.
It should be emphasized that the absence of the IR
renormalon induced power corrections obviously does
not mean the absence of the power corrections at all. It merely implies that
the relations between the ambiguities in the perturbative resummation procedures
and the true nonperturbative physics are not so evident and should be studied in
more detail.

\section{Conclusion}

The structure of the nonperturbative corrections to the quark
form factor at large momentum transfer was analyzed. In this work,
the quarks were assumed to be {\it on-}mass shell. In order to model
the nonperturbative effects, the quark scattering process was considered
in the background of the instanton vacuum.
The instanton induced contribution to the electromagnetic
quark form factor is calculated in
the large momentum transfer regime. It was shown that the
instanton induced corrections
correspond to the leading term proportional to $\ln q^2$.
The magnitude of these corrections is determined by the small instanton
liquid packing fraction parameter,
and they can be treated as finite renormalization of the subleading
logarithmic perturbative
part (\ref{eq:final}).

We have to comment that the weak-field limit used in the instanton
calculations
may deviate from the exact result. Nevertheless, we expect that using
of the instanton
solution in the singular gauge, that concentrate the field at small
distances,
leads to the reasonable numerical estimate of the full effect.
Thus, the resulting diminishing of the instanton contributions
with respect
to the perturbative result appears to be reasonable output.
It should be emphasized that
in the present paper, all the calculations have been performed
analytically while the evaluation
of the instanton contribution beyond the weak-field
approximation requires the numerical
analysis, what will be the subject of a separate work.
Besides this, the use of the singular gauge for
the instanton solution allows one to prove the exponentiation theorem for the
Wilson loop in
the instanton field \cc{DCH1} which permits to express the full
instanton contribution as the exponent
of the all-order single instanton result (\ref{eq:exp}).

It is also important to note that the results are quite sensitive
to the way one makes the integration
over instanton sizes finite. For example, if one used the sharp cutoff then the instanton
would produce strongly suppressed power corrections like $\propto (1/q)^{\beta_0}$.
However, we think that the distribution function (\ref{dist1}) should be considered
as more realistic, since it reflects more properly the structure of
the instanton ensemble modeling the QCD vacuum.
Indeed, this shape of distribution was recently advocated in \cc{SH2, DEMM99}
(see also \cc{DREV} and references therein)
and supported by the lattice calculations \cc{LAT} (for comparison, see, however,
\cc{UKQCD, SCH2}).

The instanton induced effects are more interesting for investigation and more important
for phenomenology in the hadronic processes which possess two energy scales, such as the total center-of-mass energy
$s$ (hard characteristic scale), and the squared momentum transfer $-t$ which is small compared
to the latter: $-t << s$. One of the most interesting examples of such processes is the parton-parton scattering
and the soft Pomeron problem \cc{Sh, KKL, KRP}. Another important situations where the nonperturbative
(including instanton induced) effects can emerge are
the transverse momentum distribution
of vector bosons in the Drell-Yan process (see, {\it e.g.,} \cc{KRREN}),
and the phenomenon of saturation
in deep-inelastic scattering (DIS) at small-x \cc{SCHUT}. The explicit evaluation
of the instanton effects
in some of these processes will be the subject of our forthcoming study.

\section{Acknowledgements}
\noindent

A major part of the results have been obtained in collaboration with
A.E. Dorokhov. I'm grateful to him
for numerous fruitful discussions and critical
reading of the manuscript. The useful discussions
and critical comments of N.I. Kochelev, S.V. Mikhailov and
N.G. Stefanis, as well as helpful information and enlightening
remarks of B.I. Ermolaev and L. Magnea are thanked.
The hospitality and financial support of the Organizers
of XXXI ITEP Winter School of Physics is gratefully acknowledged.
The work is partially supported by RFBR (Grant nos. 03-02-17291, 02-02-16194,
01-02-16431), Russian Federation President's Grant no. 1450-2003-2,
and INTAS (Grant no. 00-00-366).

\end{document}